\documentclass[nofootinbib,twocolumn,showpacs,preprintnumbers,pre,aps]{revtex4-1}
\usepackage{graphicx}% Include figure files
\usepackage{bm}% bold math
\usepackage{amsmath}
\usepackage{amssymb}
\usepackage{color}
\begin{document}

\title{Three-disk microswimmer in a supported fluid membrane}%

\author{Yui Ota}
\affiliation{
Department of Chemistry, Graduate School of Science and Engineering,
Tokyo Metropolitan University, Tokyo 192-0397, Japan}

\author{Yuto Hosaka}
\affiliation{
Department of Chemistry, Graduate School of Science and Engineering,
Tokyo Metropolitan University, Tokyo 192-0397, Japan}

\author{Kento Yasuda}
\affiliation{
Department of Chemistry, Graduate School of Science and Engineering,
Tokyo Metropolitan University, Tokyo 192-0397, Japan}

\author{Shigeyuki Komura}\email{komura@tmu.ac.jp}
\affiliation{
Department of Chemistry, Graduate School of Science and Engineering,
Tokyo Metropolitan University, Tokyo 192-0397, Japan}

\date{\today}
%\date{January 5, 2017}% It is always \today, today,

\begin{abstract}
A model of three-disk micromachine swimming in a quasi two-dimensional supported membrane is 
proposed. 
We calculate the average swimming velocity as a function of the disk size and the arm length. 
Due to the presence of the hydrodynamic screening length in the quasi two-dimensional fluid, 
the geometric factor appearing in the average velocity exhibits three different asymptotic behaviors
depending on the microswimmer size and the hydrodynamic screening length.
This is in sharp contrast with a microswimmer in a three-dimensional bulk fluid that shows only 
a single scaling behavior. 
We also find that the maximum velocity is obtained when the disks are equal-sized, whereas 
it is minimized when the average arm lengths are identical.
The intrinsic drag of the disks on the substrate does not alter the scaling behaviors of the geometric factor.
\end{abstract}

\maketitle
%\baselineskip=18pt

%%%%%%%%%%%%
\section{Introduction}
%%%%%%%%%%%%
\label{introduction}

Biological membranes are composed of lipid molecules and various types of proteins
which can move laterally due to the membrane fluidity~\cite{Sin72}. 
Hence biomembranes play important roles in various life processes, such as the 
transportation of materials or the reaction between chemical species~\cite{AlbertsBook}.
While some proteins are subjected to thermal agitations of lipid molecules and undergo 
passive Brownian motions~\cite{Saf75,Saf76}, there is also a large number of active 
proteins which cyclically change their conformations~\cite{Togashi2007}.
For instance, with a supply of adenosine triphosphate (ATP), some proteins act as ion 
pumps by changing their structural conformations in order to allow materials to
pass through the membranes~\cite{MBLP99,MBRP01,FLPJPB05}.
In general, such cyclic motions of proteins can lead to their active locomotion 
under certain conditions rather than just a passive motion.

By transforming chemical energy into mechanical work, microswimmers change their 
shape and move in viscous environments~\cite{Lauga09}.
Over the length scale of microswimmers, the fluid forces acting on them are governed by the 
effect of viscous dissipation.
According to Purcell's scallop theorem~\cite{Purcell77}, time-reversal body motion 
cannot be used for locomotion in a Newtonian fluid~\cite{Lauga11}.
As one of the simplest models exhibiting broken time-reversal symmetry in a 
three-dimensional (3D) fluid, Najafi and Golestanian proposed a 
three-sphere swimmer~\cite{Golestanian04,Golestanian08}, 
in which three in-line spheres are linked by two arms of varying length.
This model is suitable for analytical treatment because it is sufficient to consider only the 
translational motion, and the tensorial structure of the fluid motion can be neglected.
Recently, such a three-sphere swimmer has been experimentally realized~\cite{Grosjean2016,Grosjean2017}.
Moreover, some authors proposed a generalized three-sphere microswimmer in which the 
spheres are connected by two elastic springs with varying natural 
lengths~\cite{Pande2017,Yasuda2017}.

Compared to microswimmers in 3D bulk fluids, those in two-dimensional (2D) or quasi-2D 
fluids such as biomembranes have been less investigated in spite of their importances. 
Huang \textit{et al.}\ considered a model of an active inclusion in a membrane with three 
particles (domains) connected by variable elastic springs~\cite{Huang12}.
In their model, the natural lengths of the springs depend on the discrete states that are 
cyclically switched. 
They also performed a microscopic dynamical simulation, where the lipid bilayer structure of 
the membrane is resolved and the solvent effects are included by multiparticle collision dynamics. 
For quasi-2D fluids, there exists a hydrodynamic screening length which distinguishes 2D and 
3D hydrodynamic interactions~\cite{KomuraBook,Komura14}.
In the model by Huang \textit{et al.}, the longitudinal coupling mobility has a logarithmic 
dependence on the distance between two particles, which is valid only when the distance 
is much smaller than the hydrodynamic screening length.   
As for the mobility of a particle, they employed the 3D Stokes law even in a 2D fluid membrane, 
which is justified only when the particle size is much larger than the hydrodynamic screening length.

\begin{figure}[tbh]
\begin{center}
\includegraphics[scale=0.35]{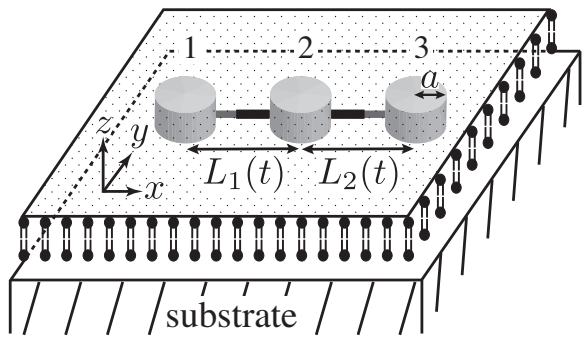}
\end{center}
\caption{A 2D three-disk micromachine swimming in a fluid membrane supported by a solid substrate.    
The flat fluid membrane located at $z = 0$ is infinitely large and its 2D viscosity is $\eta$.
The 2D momentum within the membrane can leak away due to the friction between the membrane 
and the substrate [see Eq.~(\ref{2Dhydroeq})], and hence the membrane corresponds to a 
quasi-2D fluid.
The 2D microswimmer consists of three rigid disks of radii $a_i$ ($i=1,2,3$) that are 
connected by two arms of variable lengths $L_j$ ($j=1,2$).
Without loss of generality, the microswimmer is assumed to move along the $x$-axis.
In the present model, the two arms can open and close in a prescribed form.
}
\label{membrane}
\end{figure}

In this paper, we present a systematic and also analytical investigation on the locomotion of a 
2D microswimmer immersed in a supported fluid membrane, i.e., a lipid bilayer membrane 
located on a solid substrate~\cite{Sackmann96,Tan05}.
For supported membranes, the membrane-substrate distance is usually not large, and such 
a direct contact leads to a frictional coupling between the membrane and the solid support.
Our swimmer consists of three thin rigid disks (rather than spheres) connected by two arms or 
springs which can undergo prescribed cyclic motions.
We employ the 2D mobility and the coupling mobility that take into account the 
hydrodynamic interactions mediated by the quasi-2D fluid in the presence of the 
substrate~\cite{KomuraBook}.
We analytically obtain the average velocity of such a three-disk micromachine as a function 
of the disk and arm sizes.
Due to the presence of the hydrodynamic screening length associated with the quasi-2D fluid 
model, the geometric factor in the average velocity exhibits various asymptotic size dependencies, 
which is in sharp contrast to a microswimmer in a 3D bulk fluid because they do not have any 
characteristic length scale.

In the next section, we briefly review the mobilities in a quasi-2D fluid describing a supported 
membrane. 
In Sec.~\ref{three-disk}, we discuss the motion of a 2D three-disk microswimmer in a 
supported membrane.
In Sec.~\ref{geometricfactor}, we argue various asymptotic behaviors of the geometric factor 
appearing in the average velocity.
We also examine the effects of structural asymmetry of 2D and 3D microswimmers in 
Sec.~\ref{asymmetric}.
Finally, the summary of our work and some discussions are given in Sec.~\ref{summary}.

%%%%%%%%%%%%%%%%%%%%
\section{Mobilities in a quasi-2D fluid}
%%%%%%%%%%%%%%%%%%%%
\label{quasi2D}

We first describe the quasi-2D hydrodynamic model for a supported fluid membrane as shown 
in Fig.~\ref{membrane}.
Although a lipid bilayer membrane itself can be treated as a 2D fluid, it is not an isolated 
system because the membrane is supported by the outer solid substrate (hence a 
quasi-2D fluid). 
Due to the friction between the 2D membrane and the substrate, the momentum within the 
membrane can leak away from the membrane. 
Such an effect can be taken into account through a momentum decay term in the hydrodynamic equation.
Within the Stokes approximation and by assuming the steady sate, we consider the following 2D equation
for a supported membrane~\cite{Evans88,Seki93,Ram2010}:
\begin{align}
\eta\nabla^2 \mathbf v -\nabla p-\lambda \mathbf v =0. 
\label{2Dhydroeq}
\end{align}
In the above,  $\nabla=(\partial_x, \partial_y)$ is a 2D differential operator, $\mathbf v$ (m/s) and 
$p$ (N/m) are the 2D velocity and pressure, respectively, $\eta$ (N$\cdot$s/m) is the membrane 2D viscosity, 
and $\lambda$ (N$\cdot$s/m$^3$) is the momentum decay parameter (or the friction coefficient).
In addition, we employ the 2D incompressibility condition as expressed by 
\begin{align}
\nabla \cdot \mathbf v =0.
\label{incompressible}
\end{align}

It is worthwhile to briefly mention here the physical meaning of the friction parameter $\lambda$.
For a supported membrane, a thin lubricating layer of bulk solvent [thickness $h$ and 3D 
viscosity $\eta_{\rm s}$ (N$\cdot$s/m$^2$)] exists between the membrane and the substrate~\cite{Evans88}. 
In such a situation, the friction parameter in Eq.~(\ref{2Dhydroeq}) can be identified  as 
$\lambda=\eta_{\rm s}/h$, provided that $h$ is small enough~\cite{Ram2010}.

Solving the above quasi-2D hydrodynamic equations, one can obtain the translational mobility 
coefficient $\mu$ of a rigid disk that is defined by $V=\mu F$, where $V$ is the disk velocity 
and $F$ is a driving force. 
By using the no-slip boundary condition, the resulting expression becomes~\cite{Evans88,Seki93,Ram2010}
\begin{align}
\mu(a)=\frac{1}{4\pi\eta} \left[\frac{(\kappa a )^2}{4} + 
\frac{\kappa a K_{1}(\kappa a)}{K_{0}(\kappa a)}\right]^{-1}, 
\label{1mobility}
\end{align}
where $a$ is the disk radius, $\kappa = \sqrt{\lambda/\eta}$, and $K_0$ and $K_1$ are modified 
Bessel functions of the second kind, order zero and one, respectively. 
Physically, $\kappa^{-1}$ represents the hydrodynamic screening length beyond which the
2D hydrodynamic interaction becomes irrelevant.
Notice that $\kappa^{-1}$ diverges as $\lambda \rightarrow 0$, which alludes the Stokes 
paradox in a pure 2D fluid~\cite{Saf75,Saf76}.
The effect of intrinsic drag of the disk on the substrate will be discussed later in the final section.

For $\kappa a \ll 1$, the above disk mobility asymptotically behaves as 
\begin{align}
\mu(a) \approx \frac{1}{4\pi\eta}
\left[ \ln \left( \frac{2}{\kappa a} \right) - \gamma \right],
\label{mu(a)small}
\end{align}
where $\gamma = 0.5772 \cdots$ is Euler's constant.
Here the mobility is only weakly (logarithmically) dependent on the disk size $a$, which 
is characteristic for 2D fluids.
For $\kappa a \gg 1$, on the other hand, we have
\begin{align}
\mu(a) \approx \frac{1}{\pi\eta (\kappa a)^2},
\label{mu(a)large}
\end{align}
which shows a stronger algebraic size dependence when compared with the Stokes law for 3D fluids.
Such a size dependence can be understood in terms of mass conservation principle because  
2D momentum is not conserved any more in the quasi-2D hydrodynamics~\cite{Diamant09}.

Next we explain the hydrodynamic interaction between the two disks immersed in the membrane 
by using the velocity Green's function $G_{\alpha\beta}(\mathbf r)$. 
This tensor gives the flow velocity $\mathbf v(\mathbf r)$ of the membrane at $\mathbf r$ 
due to a point force $\mathbf F$ exerted on the membrane at the origin in the $xy$-plane, 
according to $v_\alpha(\mathbf r) =G_{\alpha\beta}(\mathbf r) F_\beta$ with $\alpha,\beta=x,y$.
The velocity Green's function can generally be expressed as 
$G_{\alpha\beta}(\mathbf r)=C_1 (r) \delta_{\alpha\beta}
+C_2 (r) (r_\alpha r_\beta/r^2)$ where $\delta_{\alpha\beta}$ is the Kronecker delta and 
$r=\vert \mathbf r \vert$.
The longitudinal coupling mobility between the two disks in the membrane can be obtained from 
$M(r)=C_1(r)+C_2(r)$, where $r$ here denotes the distance between the two disks and should 
satisfy the condition $r \gg a$.
Hence $M$ does not depend on $a$ up to the lowest order contribution.

Using the quasi-2D hydrodynamic equations, the longitudinal coupling mobility $M$ is given 
by~\cite{Ramachandran10,Oppenheimer10,Ramachandran11,Komura12,Hosaka17}
\begin{align}
M(r)=\frac{1}{2\pi\eta}\left[\frac{1}{(\kappa r)^2}-\frac{K_1(\kappa r)}{\kappa r}\right].
\label{2mobility}
\end{align}
For $\kappa r \ll 1$, the above coupling mobility asymptotically behaves as 
\begin{align}
M(r) \approx \frac{1}{4\pi\eta}\left[ \ln \left( \frac{2}{\kappa r} \right) - \gamma 
+ \frac{1}{2} \right].
\label{M(L)small}
\end{align}
For $\kappa r \gg 1$, on the other hand, we have 
\begin{align}
M(r) \approx \frac{1}{2\pi\eta (\kappa r)^2}.
\label{M(L)large}
\end{align}
Equations (\ref{M(L)small}) and (\ref{M(L)large}) are analogous to Eqs.~(\ref{mu(a)small}) 
and (\ref{mu(a)large}), respectively, and the physical origins are exactly the same as 
those for the disk mobility $\mu$.

%%%%%%%%%%%%%%%%%%%
\section{Three-disk microswimmer}
%%%%%%%%%%%%%%%%%%%
\label{three-disk}

Having explained the quasi-2D hydrodynamic model for a supported membrane 
and the resulting mobilities for inclusions, we now investigate the locomotion of a microswimmer in a membrane.
To calculate the swimming velocity, we follow the procedure in Ref.~\cite{Golestanian08} for 
a three-sphere swimmer
in a 3D bulk fluid.
As shown in Fig.~\ref{membrane}, we consider a 2D micromachine consisting of three rigid
disks of radii $a_i$ ($i=1,2,3$) that are connected by two arms of variable lengths $L_j$ ($j=1,2$).
Such a three-disk microswimmer is immersed in an infinitely large and flat supported membrane having 
2D viscosity $\eta$ and the friction coefficient $\lambda$, as described before.
Each disk exerts a force $F_i$ on the quasi-2D fluid that we assume to be along the swimmer axis.
Without loss of generality, the microswimmer is assumed to move along the $x$-axis. 
In the limit $a_i/L_j \ll 1$, we can use Eqs.~(\ref{1mobility}) and (\ref{2mobility}) to 
relate the forces $F_i$ and the velocities $V_i$ as 
\begin{align}
\label{v1}
V_1& =\mu(a_1) F_1+ M(L_1)F_2+M(L_1+L_2)F_3, 
\\ 
\label{v2}
V_2&=M(L_1) F_1+\mu(a_2) F_2+M(L_2)F_3,
\\
\label{v3}
V_3&=M(L_1+L_2)F_1+M(L_2)F_2+\mu(a_3)F_3.
\end{align}

The swimming velocity of the whole object is obtained by averaging the velocities of the 
three disks, i.e., $V=(V_1+ V_2+ V_3)/3$.
Since we are interested in the autonomous net locomotion of the swimmer, there are no 
external forces acting on the disks. 
This leads to the following force-free condition:
\begin{align}
F_1+ F_2+ F_3=0.
\label{forcefree}
\end{align}
As assumed in Ref.~\cite{Golestanian08}, the motion of the arms connecting the three disks
is prescribed by the two given functions $L_j(t)$.
In this situation, the arm motions are related to the velocities as
\begin{align}
\dot{L}_1=V_2-V_1,~~~~~\dot{L}_2=V_3-V_2,
\label{armmotion}
\end{align}
where the dot indicates the time derivative.
The set of six equations in Eqs.~(\ref{v1}), (\ref{v2}), (\ref{v3}), (\ref{forcefree}), and 
(\ref{armmotion}) is sufficient to solve for the six unknown quantities $V_i$ and $F_i$ ($i=1,2,3$).

We further assume that arm deformations are relatively small as given by 
\begin{align}
L_1(t)=\ell_1+u_1(t),~~~~~L_2(t)=\ell_2+u_2(t),
\label{L1L2}
\end{align}
where $\ell_j$ are constants and $u_j/\ell_j \ll 1$.
With these prescribed arm motions, we perform an expansion of the swimming velocity
to the leading order in both $a_i/\ell_j$ and $u_j/\ell_j$.
After some calculations, we finally obtain the average swimming velocity as~\cite{Golestanian08}
\begin{align}
\overline{V}=\frac{G}{2} \langle u_1\dot{u}_2-\dot{u}_1u_2 \rangle,
\label{Vbar}
\end{align}
where $G$ is the geometric factor to be presented later in Eq.~(\ref{generalG}), 
and the averaging $\langle \cdots \rangle$ should be performed by time integration in a full cycle.
In the above calculation, the terms proportional to $u_1\dot{u}_1$, $u_2\dot{u}_2$,  and 
$u_1\dot{u}_2+\dot{u}_1u_2$ are omitted because they average out to zero in a cycle.

As studied for a three-sphere swimmer~\cite{Golestanian08}, one can assume, for example, 
that the two arms undergo the following periodic motions:
\begin{align}
u_1(t)=d_1\cos(\Omega t),~~~~~
u_2(t)=d_2\cos(\Omega t-\phi).
\end{align}
Here, $d_1$ and $d_2$ are the amplitudes of the oscillatory motions, $\Omega$ is a common 
arm frequency, and $\phi$ is a mismatch in phases between the two arms.
Then the average swimming velocity in Eq.~(\ref{Vbar}) further reads
\begin{align}
\overline{V}=\frac{G}{2} d_1 d_2 \Omega \sin \phi,
\label{barV}
\end{align}
which is maximized when $\phi=\pi/2$. 
When the disks are connected by elastic springs with time-dependent natural lengths,
a more general expression for $\overline{V}$ can be obtained~\cite{Yasuda2017}.

%%%%%%%%%%%%%%%
\section{Geometric factor}
%%%%%%%%%%%%%%%
\label{geometricfactor}

The geometric factor $G$ (having the dimension of inverse length) in Eq.~(\ref{Vbar}) 
or Eq.~(\ref{barV}) for a three-disk swimmer in a quasi-2D fluid turns out to be 
\begin{widetext}
\begin{align}
& \frac{G(\epsilon_i,\delta_j)}{\kappa} =
\nonumber \\
& 
\frac{\epsilon_1\epsilon_2\epsilon_3K_0(\epsilon_1)K_0(\epsilon_2)K_0(\epsilon_3)[\epsilon_1K_0(\epsilon_1)+4K_1(\epsilon_1)][\epsilon_2 K_0(\epsilon_2)+4K_1(\epsilon_2)][\epsilon_3K_0(\epsilon_3)+4K_1(\epsilon_3)]}{\left[4\epsilon_1K_1(\epsilon_1)K_0(\epsilon_2)K_0(\epsilon_3)+4\epsilon_2K_0(\epsilon_1)K_1(\epsilon_2)K_0(\epsilon_3)+4\epsilon_3K_0(\epsilon_1)K_0(\epsilon_2)K_1(\epsilon_3)+(\epsilon_1^2+\epsilon_2^2+\epsilon_3^2)K_0(\epsilon_1)K_0(\epsilon_2)K_0(\epsilon_3)\right]^2}
\nonumber\\
& \times
\left[2 \left(\frac{1}{\delta_1^3}+\frac{1}{\delta_2^3}-\frac{1}{\left( \delta_1+\delta_2\right)^3}\right)
- \left( \frac{K_2(\delta_1)}{\delta_1}+\frac{K_2(\delta_2)}{\delta_2}-
\frac{K_2\left(\delta_1+\delta_2\right) }{\delta_1+\delta_2} \right) \right],
\label{generalG}
\end{align}
\end{widetext}
where $\epsilon_i=\kappa a_i$ and $\delta_j=\kappa \ell_j$, and $K_2$ is modified Bessel 
function of the second kind, order two. 
We note that Eq.~(\ref{generalG}) is invariant under the exchange of not only the three disks 
$a_i$, but also under the exchange of the two arms $\ell_j$.

\begin{figure}[bth]
\begin{center}
\includegraphics[scale=0.35]{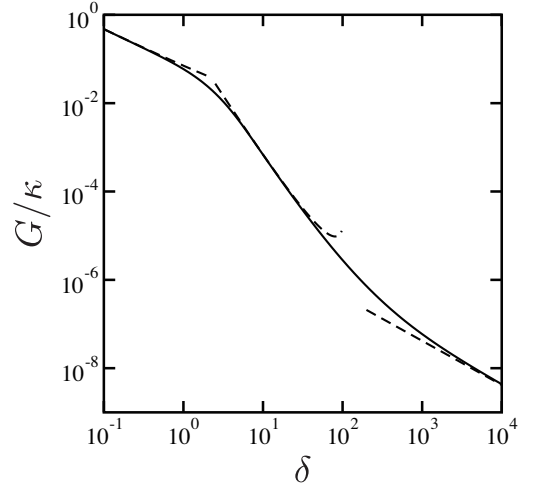}
\end{center}
\caption{
Plot of the scaled geometric factor $G/\kappa$ [see Eq.~(\ref{G(a,l)})] for a fully symmetric
2D microswimmer as a function of the scaled arm length $\delta=\kappa \ell$ 
when the disk-to-arm ratio is $a/\ell = 10^{-2}$. 
The three asymptotic expressions in Eqs.~(\ref{asymp1}), (\ref{asymp2}), and (\ref{asymp3}) 
are plotted by the dashed lines.
}
\label{Gplot}
\end{figure}

For the fully symmetric case with $a_1=a_2=a_3=a$ and $\ell_1=\ell_2=\ell$, 
the geometric factor in Eq.~(\ref{generalG})
reduces to
\begin{align}
\frac{G(\epsilon,\delta)}{\kappa} & =\frac{1}{36}
\left[\epsilon^2 + \frac{4 \epsilon K_1(\epsilon)}{K_0(\epsilon)}\right]
\nonumber \\
& \times \left[\frac{15}{\delta^3}-\frac{8K_2\left(\delta\right)}{\delta}
+\frac{2K_2\left(2\delta\right)}{\delta}\right],
\label{G(a,l)}
\end{align}
where $\epsilon=\kappa a$ and $\delta=\kappa \ell$.
Equations~(\ref{generalG}) and (\ref{G(a,l)}) are the main results of this paper.
In Fig.~\ref{Gplot}, we plot $G/\kappa$ in Eq.~(\ref{G(a,l)}) as a function of $\delta$ while 
keeping the ratio to $a/\ell=10^{-2}$.
In fact, Eq.~(\ref{G(a,l)}) has three asymptotic expressions
\begin{align}
\frac{G(\epsilon,\delta)}{\kappa} \approx \frac{1}{3 \delta [\ln (2/\epsilon)-\gamma]}
\label{asymp1}
\end{align}
for $\epsilon \ll1$ and $\delta \ll 1$, 
\begin{align}
\frac{G(\epsilon,\delta)}{\kappa}\approx\frac{5}{3 \delta^3 [\ln (2/\epsilon)-\gamma]}
\label{asymp2}
\end{align}
for $\epsilon \ll1$ and $\delta \gg 1$, and  
\begin{align}
\frac{G(\epsilon,\delta)}{\kappa}\approx\frac{5 \epsilon^2}{12 \delta^3}
\label{asymp3}
\end{align}
for $\epsilon \gg1$ and $\delta \gg 1$.

We note here that Eq.~(\ref{asymp1}) decays as $\delta^{-1}$, whereas Eqs.~(\ref{asymp2}) 
and (\ref{asymp3}) decay as $\delta^{-3}$.
The dependence on the disk size $\epsilon$ is only logarithmic in Eqs.~(\ref{asymp1}) and 
(\ref{asymp2}), while it is proportional to $\epsilon^2$ in Eq.~(\ref{asymp3}).
In Fig.~\ref{Gplot}, we also plot the above three asymptotic expressions by the dashed lines
when $a/\ell=10^{-2}$.
They are all in good agreement with the solid line that corresponds to the full expression of
Eq.~(\ref{G(a,l)}).
Notice that the apparent behavior of Eq.~(\ref{asymp3}) is $\delta^{-1}$ because we have fixed 
the ratio to $a/\ell=10^{-2}$ in this plot.

It is important to mention the physical meaning of the geometric factor $G$ in 
the average velocity.
Within the scaling argument, the geometric factor is generally related to the mobility coefficient 
$\mu$ and the longitudinal coupling mobility $M$ by 
\begin{align}
G(a, \ell) \sim \frac{M(\ell)}{\mu(a)\ell}.
\label{scaling}
\end{align}
This relation holds irrespective of the dimensionality of the micromachine 
and the surrounding fluid~\cite{YOK18}.
Due to the presence of the hydrodynamic screening length $\kappa^{-1}$ in a supported 
membrane, both $\mu$ and $M$ exhibit different asymptotic behaviors as shown in 
Eqs.~(\ref{mu(a)small}), (\ref{mu(a)large}) and Eqs.~(\ref{M(L)small}), (\ref{M(L)large}),
respectively.
Various limiting expressions of $G$ in Eqs.~(\ref{asymp1})-(\ref{asymp3}) can 
be understood as different combinations of the asymptotic forms of $\mu$ and $M$.
For example, Eq.~(\ref{asymp3}) showing the scaling $G \sim (a^2/\ell^2)/\ell$ is a direct
consequence of Eqs.~(\ref{mu(a)large}) and (\ref{M(L)large}).
We also note that, because of the explicit $1/\ell$-dependence in Eq.~(\ref{scaling}), 
the logarithmic dependence of $M(\ell)$ on $\ell$, as in Eq.~(\ref{M(L)small}), does
not show up in Eq.~(\ref{asymp1}) within the lowest order expansion.

\begin{figure}[tbh]
\begin{center}
\includegraphics[scale=0.35]{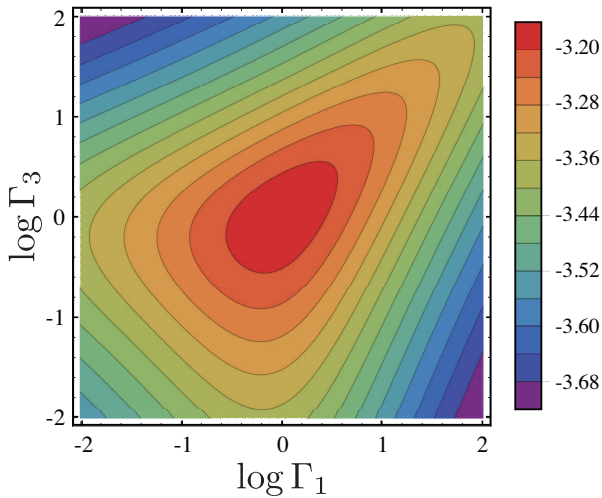}
\end{center}
\caption{
Color plot of the scaled geometric factor $\log (G/\kappa)$ [see Eq.~(\ref{generalG})] of a 2D 
microswimmer as a function of the asymmetry parameters $\Gamma_1=a_1/a_2$ and 
$\Gamma_3=a_3/a_2$ 
[see Eq.~(\ref{asympara})] when $\kappa a =0.1$  and $\kappa \ell =10$.
Notice that $a=(a_1+a_2+a_3)/3$ and $\ell=\ell_1=\ell_2$.
All the quantities including the color scale are plotted in the logarithmic scale.
The maximum of $G$ occurs at $\Gamma_1=\Gamma_3=1$. 
}
\label{asymmetry}
\end{figure}

\begin{figure}[bth]
\begin{center}
\includegraphics[scale=0.35]{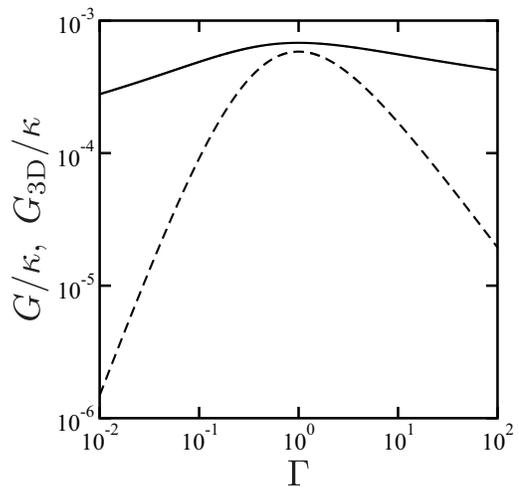}
\end{center}
\caption{
Plot of the scaled geometric factor $G/\kappa$ (solid line) [see Eq.~(\ref{generalG})] 
and $G_{\rm 3D}/\kappa$ (dashed line) [see Eq.~(\ref{G3D})] as a function of the 
asymmetry parameter $\Gamma=\Gamma_1=\Gamma_3$ when $\kappa a =0.1$ and 
$\kappa \ell =10$.
Notice that $a=(a_1+a_2+a_3)/3$ and $\ell=\ell_1=\ell_2$.
The maxima of $G$ and $G_{\rm 3D}$ occur at $\Gamma=1$.}
\label{asymdiag}
\end{figure}

In order to compare our result with that of a three-sphere swimmer in a 3D fluid, we show here 
its corresponding geometric factor as obtained in Ref.~\cite{Golestanian08}: 
\begin{align}
G_{\rm 3D}=\frac{3 a_1a_2a_3}{(a_1+a_2+a_3)^2}
\left[ \frac{1}{\ell_1^2}+\frac{1}{\ell_2^2}-
\frac{1}{(\ell_1+\ell_2)^2} \right],
\label{G3D}
\end{align}
where $a_i$ here denote the radii of the three spheres (rather than disks).
For the symmetric case with $a_1=a_2=a_3=a$ and $\ell_1=\ell_2=\ell$, the above 
expression reduces to
\begin{align}
G_{\rm 3D}=\frac{7a}{12 \ell^2}.
\label{G3Dsym}
\end{align}
First, we note that Eq.~(\ref{G3D}) or Eq.~(\ref{G3Dsym}) does not depend on the 3D fluid viscosity, 
while Eq.~(\ref{generalG}) or Eq.~(\ref{G(a,l)}) is dependent on the membrane viscosity 
$\eta$ through the inverse screening length $\kappa$. 
Second, the essential size dependence in Eq.~(\ref{G3Dsym}) is 
$G_{\rm 3D} \sim (a/\ell)/\ell$ which does not appear in the previous quasi-2D case.
On the other hand, such a dependence is in accordance with the scaling relation 
Eq.~(\ref{scaling}) and the Stokes law in a 3D fluid without any characteristic length scale.
Hence the existence of the characteristic length scale, $\kappa^{-1}$, for a quasi-2D fluid 
completely changes the asymptotic size dependencies of the average velocity.
This is an important finding of this paper and highlights the essential difference between 
2D and 3D microswimmers.

%%%%%%%%%%%%%%%%%%%%
\section{Asymmetric microswimmers}
%%%%%%%%%%%%%%%%%%%%
\label{asymmetric}

Since we have obtained the general expression of the geometric factor $G$ for a three-disk 
microswimmer, as shown in Eq.~(\ref{generalG}),  we discuss now the effects of structural
asymmetry of a microswimmer on its geometric factor.  
We first set as $\ell_1=\ell_2=\ell$ and vary the two ratios between the disk sizes as defined by
\begin{align}
\Gamma_1=\frac{a_1}{a_2},~~~~~
\Gamma_3=\frac{a_3}{a_2},
\label{asympara}
\end{align}
whereas we keep, for instance, the sum of the three radii being fixed to $a_1+a_2+a_3=3a$.
In Fig.~\ref{asymmetry}, we plot the scaled geometric factor $G/\kappa$ in 
Eq.~(\ref{generalG}) as a function of the two ratios $\Gamma_1$ and $\Gamma_3$
when $\kappa a =0.1$ and $\kappa \ell =10$.
Notice that $a/\ell$ should be small within our expansion scheme and the color scale 
indicates the quantity $\log (G/\kappa)$.
From this plot, we find that the maximum of the geometric factor is realized when the 
disk size is identical, i.e., $\Gamma_1=\Gamma_3=1$.
In other words, any asymmetry in the disk size leads to a reduction of the average swimming 
velocity.

\begin{figure}[bth]
\begin{center}
\includegraphics[scale=0.35]{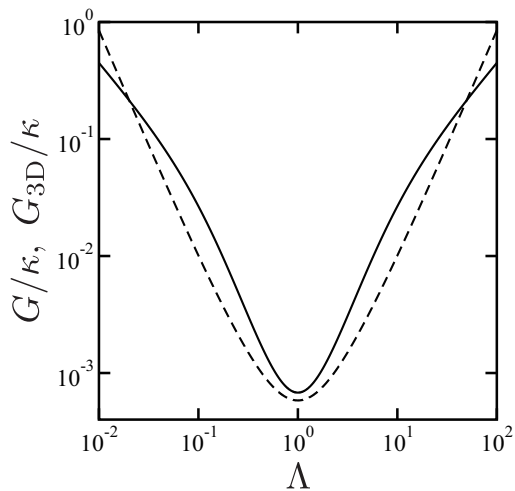}
\end{center}
\caption{
Plot of the scaled geometric factor $G/\kappa$ (solid line) [see Eq.~(\ref{generalG})] 
and $G_{\rm 3D}/\kappa$ (dashed line) [see Eq.~(\ref{G3D})] as a function of the 
asymmetry parameter $\Lambda=\ell_1/\ell_2$ [see Eq.~(\ref{Lambda})] when 
$\kappa a =0.1$ and $\kappa \ell =10$.
Notice that $a=a_1=a_2=a_3$ and $\ell=(\ell_1+\ell_2)/2$.
The minima of $G$ and $G_{\rm 3D}$ occur at $\Lambda=1$.
}
\label{asymmetry2}
\end{figure}

In Fig.~\ref{asymdiag}, we further consider the case when $\Gamma_1=\Gamma_3=\Gamma$ 
(or, equivalently, $a_1=a_3$), and plot $G/\kappa$ as a function of $\Gamma$. 
Such a plot corresponds to a cross-section of Fig.~\ref{asymmetry} along the diagonal line.
To compare the quasi-2D case with the 3D case, we also plot $G_{\rm 3D}/\kappa$ [see 
Eq.~(\ref{G3D})] as a function of $\Gamma$ under the same condition. 
For $G_{\rm 3D}$, the ratios $\Gamma_1$ and $\Gamma_3$ correspond to those between 
the sphere radii.
It should be also noted that Eq.~(\ref{G3D}) does not contain $\kappa$ and it is used only 
to compare with the quasi-2D case. 
Although both $G$ and $G_{\rm 3D}$ are maximized at $\Gamma=1$, the dependence 
on $\Gamma$ is much weaker for $G$.  
This weak dependence originates from the logarithmic dependence of the disk mobility 
$\mu$ on the disk size $a$, as shown in Eq.~(\ref{mu(a)small}).

Alternatively, one can set as $a_1=a_2=a_3=a$ and vary the ratio between the arm 
lengths defined by
\begin{align}
\Lambda=\frac{\ell_1}{\ell_2},
\label{Lambda}
\end{align}
whereas we keep the sum of the two arm lengths being fixed to $\ell_1+\ell_2=2\ell$.
In Fig.~\ref{asymmetry2}, we plot the scaled geometric factors $G/\kappa$ and 
$G_{\rm 3D}/\kappa$ as a function of $\Lambda$ when $\kappa a =0.1$ and $\kappa \ell =10$ 
as before.
Here it is remarkable that both $G$ and $G_{\rm 3D}$ are minimized when $\Lambda=1$.
On the other hand, the overall behaviors of $G$ and $G_{\rm 3D}$ are rather similar because the
chosen parameter value satisfies the condition $\kappa \ell \gg 1$ for which the coupling mobility 
$M$ exhibits an algebraic dependence on the distance $\ell$, as shown in Eq.~(\ref{M(L)large}).

%%%%%%%%%%%%%%%%%%
\section{Summary and discussion}
%%%%%%%%%%%%%%%%%%
\label{summary}

In summary, we have proposed a model of 2D three-disk micromachine swimming in a quasi-2D 
supported membrane. 
In particular, we have obtained the average swimming velocity as a function of the disk size 
and the arm length. 
Due to the presence of the hydrodynamic screening length in the quasi-2D fluid, 
$\kappa^{-1}$, the geometric factor in the average velocity exhibits various asymptotic 
behaviors depending on the microswimmer size and the screening length.
Our result has been confirmed by the scaling argument for the geometric factor. 
We have also looked at the effects of structural asymmetry of a microswimmer, and found 
that the geometric factor is maximized when the disks are equal-sized, whereas it is minimized 
when the average arm lengths are identical.

At this point, a rough estimate of the characteristic length scales would be 
useful~\cite{Ram2010,Ramachandran11,Hosaka17}.
The membrane viscosity of lipid bilayers at physiological temperatures is approximately 
$\eta \approx 10^{-9}$~N$\cdot$s/m and the viscosity of surrounding water is 
$\eta_{\rm s} \approx 10^{-3}$~N$\cdot$s/m$^2$.
For supported membranes,  we can approximate the height of the intervening solvent region 
as $h \approx 10^{-9}$~m.
Hence we obtain $\kappa^{-1} = \sqrt{\eta h/\eta_{\rm s}} \approx 3 \times 10^{-8}$~m.
We note that this length scale is relatively small and the large scale behavior is expected for
micron-sized swimmers.

In our model of a three-disk microswimmer, we have assumed that the three disks are 
connected by two arms and their time-dependent motions are given by 
Eq.~(\ref{L1L2})~\cite{Golestanian04,Golestanian08}.
Alternatively, one can also consider a three-disk microswimmer in which the disks are 
connected by two elastic springs, while the natural length of each spring is assumed to 
undergo a prescribed cyclic change.
Using the results in Ref.~\cite{Yasuda2017}, one can immediately estimate the average 
quantity $\langle u_1\dot{u}_2-\dot{u}_1u_2 \rangle$ in Eq.~(\ref{Vbar}) and 
obtain the average velocity $\overline{V}$ of an elastic 2D microswimmer. 
It can be generally shown that the swimming velocity increases with frequency in the 
low-frequency region, whereas in the high-frequency region, the average velocity decreases 
when the frequency is increased~\cite{Yasuda2017}. 
Such a behavior originates from the intrinsic spring relaxation dynamics of an elastic swimmer.

Although we have taken into account the hydrodynamic friction 
between the fluid membrane and the substrate through the friction coefficient $\lambda$
in Eq.~(\ref{2Dhydroeq}), the effect of intrinsic drag of the disks on the substrate was not 
considered in Eq.~(\ref{1mobility}).
One can naturally assume that the drag coefficient of a disk on the substrate is proportional to its area and 
is given by $\lambda_{\rm d}\pi a^2$, where $\lambda_{\rm d}$ is the disk friction coefficient. 
In this case, the translational mobility coefficient in Eq.~(\ref{1mobility}) will be modified~\cite{Evans88}:
\begin{align}
\mu(a)=\frac{1}{4\pi\eta} \left[\frac{(\kappa a )^2}{4} 
\left(1+\frac{\lambda_{\rm d}}{\lambda} \right) + 
\frac{\kappa a K_{1}(\kappa a)}{K_{0}(\kappa a)}\right]^{-1}.
\label{1mobilitygen}
\end{align}
Notice that the correction term due to $\lambda_{\rm d}/\lambda$ gives rise to a 
contribution that is independent of $\lambda$.
Since only the coefficient of $(\kappa a)^2$ is altered when compared with Eq.~(\ref{1mobility}),  
the drag acting on the disks modifies our result only up to a numerical factor when $\kappa a \gg 1$.
This means that the asymptotic scaling behaviors in Sec.~\ref{geometricfactor} are not affected 
by the drag forces on the disks.
In principle, the disk friction coefficient can be different for different disks. 
Such an effect can be effectively taken into account by considering different disk radii as long as 
they are much larger than the screening length $\kappa^{-1}$.

In this paper, we have discussed the behavior of a 2D microswimmer in a quasi-2D fluid that 
is characterized by a hydrodynamic screening length, $\kappa^{-1}$.
It should be noted, however, that we encounter a similar situation in which a 3D micromachine
swims in a structured viscoelastic fluid having a characteristic length scale.
According to our preliminary result, we find that the frequency dependence of the average 
velocity exhibits fairly complex behaviors depending on the machine size relative to the 
characteristic length scale of the surrounding structured fluid.
Details of such an investigation will be reported elsewhere~\cite{YOK18}.

In the future, we shall also investigate the case when the surrounding bulk fluid is 
viscoelastic~\cite{Komura12}.
Such a study will enable us to obtain the frequency dependent complex viscosity of the 
surrounding 3D fluid by measuring the velocity of a 2D microswimmer in a 
membrane~\cite{Yasuda17SMR}.
Such a method will provide us with a new type of non-contact surface microrheology.

%%%%%%%%%%
\acknowledgments
%%%%%%%%%%

We thank  R.\ Okamoto for helpful discussions.
S.K.\ acknowledges support by Grant-in-Aid for Scientific Research on Innovative 
Areas ``\textit{Fluctuation and Structure}" (Grant No.\ 25103010) from the 
Ministry of Education, Culture, Sports, Science, and Technology (MEXT) of Japan,
and by Grant-in-Aid for Scientific Research (C) (Grant No.\ 15K05250 and 18K03567) 
from the JSPS.

%%%%%%%%%%%%%%%

\end{document}